\documentclass[a4paper,twoside]{article}
\newcommand{\affil}[1]{$^{\rm #1}$}
\usepackage[authoryear]{natbib}
\bibpunct{(}{)}{;}{a}{}{,}
\usepackage{graphicx}
\graphicspath{{converted_graphics/}} 
\date{} 
\title{\large\bf\flushleft Morphological Structures of Planetary Nebulae}
\author{\parbox{\textwidth}{\flushleft
\vspace{-0.5cm}
{\it Sun Kwok\affil{A, B}}\\
\vspace{0.4cm}
{\small \affil{A}\,Faculty of Science, The University of Hong Kong, Hong Kong, China}\\
{\small \affil{B}\,Email: sunkwok@hku.hk}}}
\begin{document}
\twocolumn[
\begin{changemargin}{.8cm}{.5cm}
\begin{minipage}{.9\textwidth}
\vspace{-1cm}
\maketitle
%
%
\small{\bf Abstract:}

Since various structural components of planetary nebulae manifest themselves differently, a combination of optical, infrared, submm, and radio techniques is needed to derive a complete picture of planetary nebulae.  The effects of projection can also make the derivation of the true 3-D structure difficult.  Using a number of examples, we show that bipolar and multipolar nebulae are much more common than usually inferred from morphological classifications of apparent structures of planetary nebulae. \\

We put forward a new hypothesis that the bipolar and multipolar lobes of PN are not regions of high-density ejected matter, but the result of ionization and illumination.  The visible bright regions are in fact volume of low densities (cleared by high-velocity outflows) where the UV photons are being channelled through.  We suggest that multipolar nebulae with similar lobe sizes are not caused by simultaneous ejection of matter in several directions, but by leakage of UV photons into those directions.


\medskip
\medskip
\end{minipage}
\end{changemargin}
]
\small

\section{Introduction}

Planetary nebulae (PNe) are traditionally identified by their morphological appearances in photographic and imaging surveys, or by their spectroscopic properties in objective prism or emission-line surveys.  In recent years, new PNe candidates have also identified through their spectral properties among objects in radio surveys, or through their colors among objects in infrared surveys.  Although a number of observational properties have been employed as selection critera for PNe, extensive confusion still exist between other classes of objects such as H{\sc ii} regions, symbiotic stars, Wolf-Rayet nebulae \citep{kwo00}.

Recent narrow-band imaging CCD surveys such as the Anglo-Australian Observatory UK Schmidt Telescope H$\alpha$ survey \citep{par05, mis08} and the Isaac Newton Telescope Photometric H$\alpha$ Survey of the Northern Galactic Plane (IPHAS) \citep{dre05} have greatly expanded the number of PNe candidates.  Comparison between the brightness in the narrow-band (H$\alpha$) and broad-band images easily identifies the emission-line objects, and the images allow for morphological identification.  The PN nature of the candidates can be further strengthened by follow-up spectroscopic observations \citep{par06}.
The expaneded list of PNe in the Macquarie/AAO/Strasbourg H$\alpha$ Planetary Nebula Catalogue (MASH) allows for a more systematic study of the morphology of PNe.  For example, the improved sensitivity of the recent surveys makes possible the detection of low-surface brightness objects, which could represent a population of evolved PNe.

\section{Morphological classification}

PNe are classified into morphological classes based on their apparent structure.  Beginning with \citet{cur18}, there have been many attempts in morphological classifications based on observational surveys of PNe \citep[e.g.,][]{khr68, sta93, aaq96, man96}.  \citet{par06} classify the 903 new PNe in the MASH catalogue into 6 classes: round, elliptical, bipolar, irregular, asymmetric and star-like.  The most basic classification is that of \citet{bal87}, who classifies PNe into classes of round, elliptical, and butterfly. The morphological classification studies have often emphasized the small fraction ($\sim$15\%) of PNe having bipolar structures.  Is this small percentage an accurate reflection of the fraction of PNe with  bipolar morphologies?
Often the true intrinsic structures of PNe cannot be easily derived from the observed images, as all classification
schemes suffer from the following problems:

\begin{itemize}

\item {Sensitivity dependence: a deeper exposure can reveal fainter
structures which change the classification of the PNe.  For example,  
the waist of a bipolar nebula could be classified as elliptical if the
bipolar lobes are too faint to be detected.  NGC 650-1\index{NGC~650-1}, Sh 1-89\index{Sh~1-89}, and SaWe 3 \index{SaWe~3}are some of the cases where their bipolar nature were only
revealed after deep CCD imaging \citep{hua97, hua98}.}

\item  {Species dependence:  the morphology of PNe observed in lines of
different ions is not necessarily the same, as the result of 
ionization structures and stratification effects.}

\item  {Projection effects: morphology classifications describe the
two-dimensional apparent structures, not the intrinsic
structures of the PNe \citep{khr68, man04}.}  
\end{itemize}

A bipolar nebula with an equatorial waist viewed at an angle other than edge on will have a ring-like appearance, especially when the bipolar lobes are faint.  The best-studied PN NGC 7027 has a very prominent ring structure, but when studied in detailed, it is revealed to be a bipolar nebula \citep{lat00}.  The same is true for NGC 3132 \citep{mon00}.
Even well-known objects such as NGC 6720 (the Ring Nebula) and NGC 7293(the Helix Nebula) turn out to have bipolar morphologies \citep{bry94, mea05a}.  Two objects with very different apparent morphologies, such as the Ring and NGC 6853 (the Dumbbell Nebula), may in fact have similar intrinsic 3-D structures \citep{kwo08}. 

\begin{figure}[tbp]
\includegraphics[width=3in,height=3in, keepaspectratio]{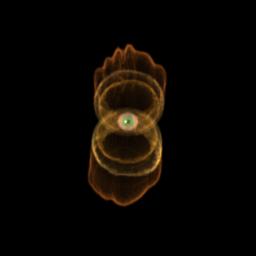} 
\includegraphics[width=3in,height=3in,keepaspectratio]{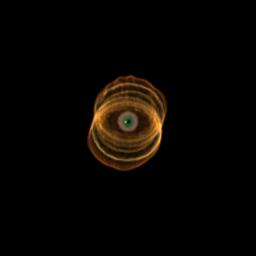} 
\caption{A SHAPE \citep{ste_lop06} model image of MyCn 18 (top) and the same model rotated to another viewing angle that resembles the Helix Nebula (bottom). The model was first created to animate the {\it HST} image of MyCn 18 \citep{sah99} and then rotated to other orientations.  Models in Fig.~1 and 2 are constructed by N. Koning.}
\label{mycn18}
\end{figure}

\begin{figure}[tbp]
\includegraphics[width=3in,height=3in,keepaspectratio]{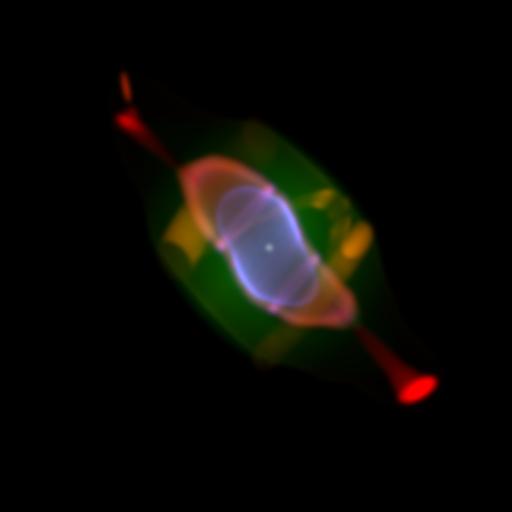} 
\includegraphics[width=3in,height=3in,keepaspectratio]{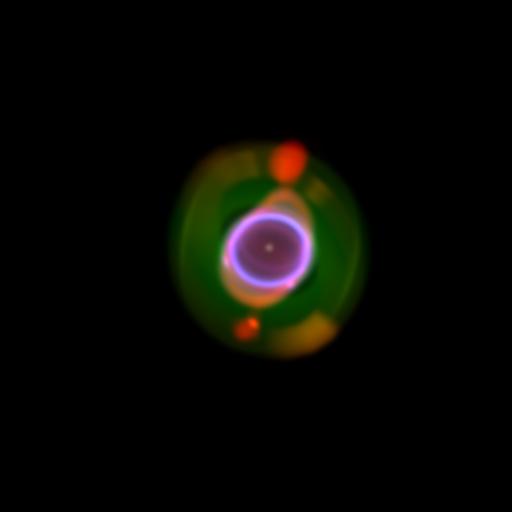} 
\caption{A SHAPE model image of NGC7009 (top) and the same model rotated to another viewing angle that resembles NGC 6826 (bottom).  The model was first created to animate the {\it HST} image of NGC 7009 \citep{bal98} and then rotated to other orientations.}
\label{ngc7009}
\end{figure}

In Figures \ref{mycn18} and \ref{ngc7009}, we show models of MyCn 18 and NGC 7009 respectively.  By rotating the model images to another orientation, we can see that same objects could resemble other well known PNe, namely NGC 7293 (the Helix Nebula) and NGC 6826, respectively.  A similar 3-D reconstruction exercise has also been done by \citet{sab04}.
These examples clearly illustrate that apparent morphology  alone is
not sufficient to obtain the true intrinsic structure of PN.  Kinematic
data are necessary to separate various components projected on the same
positions in the sky.

\section{Multi-wavelength observations of PNe}

A common perception of PN is that a shell of gas photo-ionized by a hot central star.  Now we know that a typical PN is made up of ionized, atomic, molecular, and dust components, and these components do not necessarily have the same geometry.  Their temperatures range from 100 K to 10$^6$ K and need to be studied with different observational techniques from radio to X-ray.  The optical component, the most easily observed and therefore the best studied, show multiple structures in the form of shell, crown, halo, and lobes.  These structures have been successfully modeled by 1-D dynamical evolution of PNe incorporating stellar evolution, interacting winds dynamics, and time-dependent photoionization \citep{per04, ste06}.

The existence of a dust torus is often inferred from the presence of a dark lane in the optical image of bipolar nebulae.  Mid-infrared imaging can provide a direct determination of the size and orientation of the torus \citep{16594}.   In the proto-PN IRAS 17441$-$2411 where the the orientation of the axis of the infrared torus is found to be offset by 23$^\circ$ from the bipolar axis.  If the torus is responsible for or related to the collimation of the bipolar outflow, this raises the possibility that the outflow has undergone precession over the last 100 years \citep{vol07}.
From observations obtained with the Multiband Imaging Photometer (MIPS) of the {\it Spitzer Space Telescope},  \citet {su04} shows that warm dust (as evidenced by emission at 24 $\mu$m) is present in both the bipolar lobes and in the torus of NGC 2346, cool dust (as evidenced by emission at 70 $\mu$m) is mostly located in the edge-on equatorial torus, and the cold dust (as evidenced by emission at 160 $\mu$m) is distributed over a spherical region, probably arising from the remnant of the AGB wind.

With submm arrays such as {\it PdBI, CARMA, SMA}, and {\it ALMA},  the distribution of the emission regions of molecular lines, free-free continuum, and dust continuum can be mapped with subarcsec angular resolutions.  
Comparisons between the molecular, optical, and radio continuum maps of NGC 7027 show a clear photodissociation region separating the ionized and molecular gas components \citep{gra93}.
For NGC 6302, submm observations have found an expanding molecular torus, which correspond to the dark lane separating the optical bipolar lobes \citep{per07, trung}.

Fast molecular outflows can be detected by single-dish mm/submm observations from the extended wings of the CO profile.
High-resolution CO $J=6-5$ images of AFGL 618 show that the high-velocity components lie along the optical lobes \citep{nak07}.  In NGC 2440, the CO outflow has been imaged to correspond to one pair of the optical bipolar lobes \citep{wan08}.

With the construction of {\it ALMA}, we will be able to map the dust and molecular components of PNe  with much greater sensitivity and angular resolution.  Only through a comprehensive study of the ionized gas/molecular gas/dust components we can arrive at a full picture of the morphology of PNe.

\section{Molecular Hydrogen as a Tracer of Dynamical Interactions}
\label{shock}

\begin{figure}
 \includegraphics[width=3in]{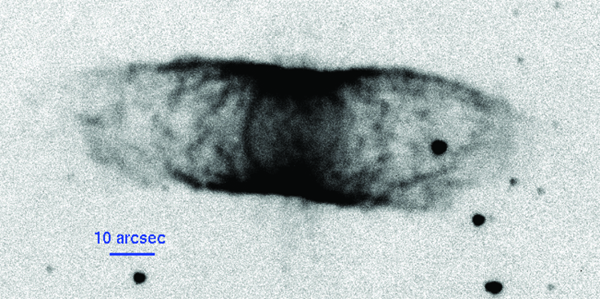}
\\
\includegraphics[width=3in]{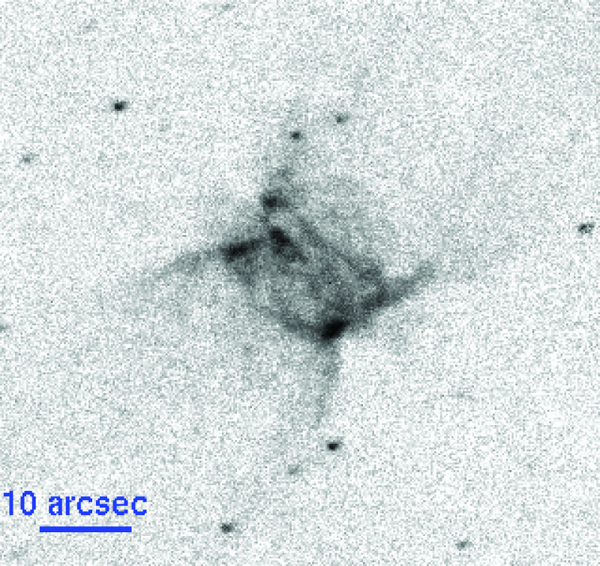}
\caption{Continuum-subtracted H$_2$ image of IC 4406 (top) and K3-72 (bottom) obtained with the Canada-France-Hawaii Telescope.  The bipolar nebulae, especially the waist of the nebulae, are clearly confined externally by unseen neutral gas.}
\label{cfht1}
\end{figure}

The 2.12 $\mu$m vibrational-rotational line of H$_2$ is a useful tracer of dynamical interactions in PNe.  
A narrow-band H$_2$ image can often reveal structures not seen in broad-band images.  Fig.~\ref{cfht1} shows H$_2$ image of two bipolar PNe IC 4406 and K3-72.  In both cases, rays of equatorial outflows can be seen.  Since the IC 4406 is inclined at a angle, a bright elliptical ring can clearly be seen in the waist region.  The bright rims in the lobes of IC 4406 sharply define the edges of the lobes, suggesting
that the lobes are confined by an external medium and that the H$_2$ emission trace regions
of wind interactions.   The H$_2$ emissions at the caps of the lobes probably represent the
regions where the fast wind is breaking through the remnant AGB circumstellar envelope.  We note that these features are much more prominent in this H$_2$ image than in the {\it HST} optical image of IC 4406 \citep{odel02}.
The equatorial features resemble the high-speed ``skirts'' seen around the waist of NGC 6302 \citep{mea05b}, or the ``chakram'' of Mz-3 \citep{san04}.  
Although K3-72 has a more edge-on orientation, similar structures of equatorial torus, bipolar lobes, and radial equatorial jets can also be seen.

We can infer from Fig.~\ref{cfht1} that the tight waists of IC 4406 and K3-72 are due to confinement by an unseen equatorial torus and the nebulae are ionization bounded in the equatorial directions.  Infrared and submm imaging will be needed to determine the geometric and dynamical relationship between the torus and the bipolar lobes.

\section{Multipolar nebulae}

\begin{figure}
\includegraphics[width=3in, angle=0]{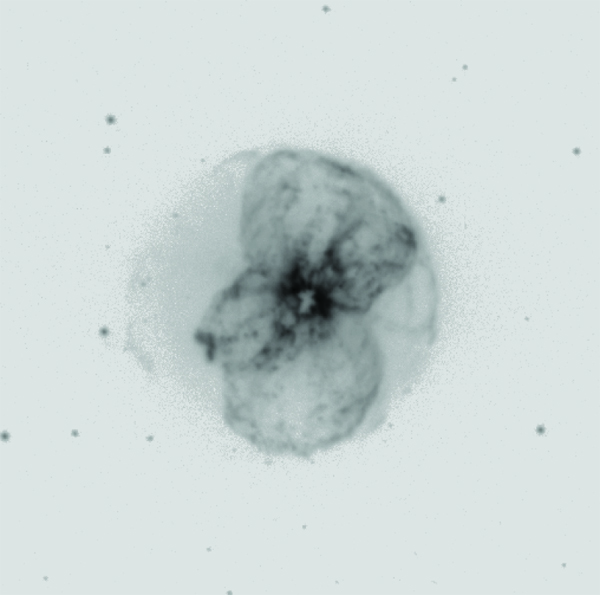}
\caption{An [N {\sc ii}] image of NGC 2440 taken at the Canada-France-Hawii Telescope showing two pairs of bipolar lobes.  The round outer shell probably represent the extent of the AGB envelope.  Are the multiple lobes manifestations of separate ejection events or photoionized regions bounded by neutral gas?  (see discussions in Section \ref{illum}).} 
\label{ngc2440}
\end{figure}

\begin{figure}
\includegraphics[width=3in, angle=0]{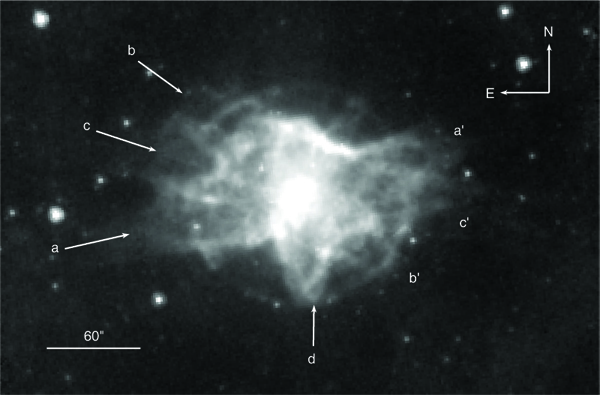}
\caption{Spitzer IRAC 8 $\mu$m image of NGC 6072 with the three possible pairs of bipolar lobes marked as $a-a^\prime, b-b^\prime, c-c^\prime$.  The equatorial disk is marked as $d$.  Are the multiple lobes regions of separate ejections or directions of UV photon leakage? \label{marked}}
\end{figure}

Although planetary nebulae are commonly believed to possess axi-symmetry, an increasing number of planetary nebulae has been found to be point-symmetric, including the class of multipolar nebulae \citep{man96b}.  The most well-known example is NGC 2440, which in addition to its bright pair of bipolar lobes, have been found to have two other fainter pairs of lobes \citep[see also Fig.~\ref{ngc2440}]{lop98}.  A survey of compact planetary nebulae by the {\it HST} has also revealed further multipolar objects such as M1-37 and He2-47 \citep{sah00} and NGC 6881 \citep{su05}.  NGC 6072, a bipolar PN with a bright waist \citep{cor95}, in fact possess two or three pairs of bipolar lobes with the bright central part being an equatorial ring (Fig.\ref{marked}).

At the present time, we may be just scratching the surface of the multipolar phenomenon.  \citet{mam06} have found a new PN with an equatorial ring, a pair of inner lobes, a pair of bright main lobes, and very large faint outer lobes.  The symmetry axes of these lobes all seem to have different orientations.  Further deep and wide-field imaging of PNe will lead to a greater appreciation of the complexity of the PN morphology.

If we model a multipolar nebula, e.g., by systems of equatorial rings plus bipolar lobes, and view the objects from different perspectives, one gets very different impressions of the object.  For example, the equatorial rings viewed near pole on will be taken as projection of a spherical shells, and the projections of the other lobes will be assumed to be internal structures of the shell.
The apparent morphologies of these different projections can easily mislead the observer about the intrinsic structure of the nebula.

\section{Effects of illumination}
\label{illum}

It is important to remember that an optical image of a PN does not give a complete representation of the distribution of matter.  We have already mentioned that there is ``dark matter'' in the form of molecular and solid-state matter in PNe, and their masses can be considerably higher than the total mass in the ionized region.  I have long argued that since PPN shine by reflected star light, the visible image that we see in a PPN in fact represent the cavities where light can escape, a complete opposite to the actual distribution of matter \citep{kwo04}.  Even in PNe, the bright optical emission region may represent low-density regions cleared out by high-velocity outflows.  In bipolar nebulae such as IC 4406  shown in Fig.~\ref{cfht1}, the optical nebula is probably ionization bounded and bipolar lobes are ionized because these are the directions where the UV photons are channeled through.  

In this model, the morphology of PNe is not so much defined by regions of matter ejection, but are defined by ``holes'' in the matter distribution where densities are low enough for the available UV flux to ionize.  We cannot have a complete description of matter distribution until we have mappings of both the ionized and neutral gas components.


\section{Dynamical evolution}

The fraction of PNe with bipolar structure has important implications on our understanding of PNe dynamical evolution.  If the real fraction of bipolar PNe is indeed high, then one is compelled to seek a physical scenario of how such a morphological transformation occurs.  From molecular-line imaging, we know that the mass-loss process is mainly spherically symmetric, with the possible departure from spherical symmetric at the very end.  In the following post-AGB evolution, observations of proto-planetary nebulae suggest that one or more highly collimated outflows develop, carving out bipolar cavities in the spherical remnant AGB envelope.  Shell structures develop as the result of the sweeping up of remnant AGB material by these fast outflows.  Although the above scenario roughly describes the morphological transformation of PNe, the details are not known.  For example, it is not clear whether an equatorial outflow develop first, forming an equatorial disk which collimates the bipolar flow, or vice versa \citep{hug07}.

An interesting aspect of the multi-polar phenomenon is that the sizes of the multi-polar lobes are approximately the same, suggesting that they originate at approximately the same time.  How a PN can simultaneously eject several pairs of highly collimated outflows is a mystery.  Most models attribute the multipolar phenomenon to precession of the collimating disk, with either binarity or magnetic as the agent.  Here we offer an alternative explanation.  According to the theory outlined in Section \ref{illum}, the multipolar lobes only reflect directions where the equatorial torus has ``holes''.  If this is the case, the explanation to the multipolar phenomenon lies not so much in the dynamical ejection, but in how such ``holes'' are created.  The similarity in sizes of the multiple lobes is therefore naturally explained, and no simultaneous ejection is required.

The MASH survey has found many low-surface brightness, round, PNe.  If they correspond to an evolved stage of evolution of PNe, one has to account for the dynamical morphological evolution from bipolar back to spherical symmetry.  This, e.g., can be attributed to the increasing dominance of the effect of thermal pressure of the bubble and the effect of recombination as the UV output of the central star declines \cite{ste06}.  If we again appeal to the theory put forward in Section \ref{illum}, an evolved PN will be entirely density bounded and the spherical morphology just reflects the original morphology of the AGB envelope.

\section{What is a PN?}

Planetary nebulae are usually defined by a combination of observation properties of the nebula by morphology (having some degree of symmetry), by spectrum (emission-line spectrum with little or no continuum, H$\alpha$ to [O~{\sc iii}] ratio, expansion velocity, etc.), and the properties of the central star (temperature, luminosity, gravity, etc.).  However, many symbiotic stars (in particular those of D-type) share similar properties.  Symbiotic stars and PNe even have similar properties in the radio, infrared, and X-ray spectral regions.  There are some distinctions, however.   Symbiotic stars have higher excitation lines as the result of accretion, periodic photometric variability due to the pulsation of the cool component, and molecular absorption features produced in the cool star's atmosphere \citep{kwo03}.  It has also been suggested that symbiotic stars have more extreme bipolar morphologies, e.g., as in Hen 2-104 and R Aqr \citep{cor03}.


When a new class of astronomical object is first discovered, they are usually first defined by their observational properties.  But as we gain insight into the phenomonon, a more precise definition is given based on the physical or evolutionary nature of the class of objects.  For PNe, we can define  them as ionized expanding circumstellar shells showing some degree of symmetry ejected by a hot, compact, central star evolving from the AGB to the white dwarf phase \citep{kwo00}.
This definition will distinguish PNe from symbiotic stars or novae, which are binary systems undergoing mass exchange.  
While the hydrogen envelope mass of the a PN central star is being depleted by nuclear burning and mass loss, and therefore constantly evolving, a symbiotic star maintains its energy source through accretion, and is therefore stationary (or even going backwards) in evolution \citep{pac80}.  It should be noted, however, that a post-outburst nova or symbiotic nova also evolves to the blue similar to a PN.  
Unless the outburst (H ignition) is observed, it may also be difficult to distinguish a symbiotic nova (e.g., V1016 Cyg and HM Sge) from a PN.  The separation of symbiotic stars from PNe in PN catalogues will be a long continuous process \citep{koh94}.  For PNe in external galaxies, this is almost an impossible task.



Since a binary system can undergo mass exchange either when the primary or the secondary is in the post-main-sequence phase of evolution, the number of evolution scenarios (depending on initial mass ratios and separations) is almost infinite \citep{web88}.  When a binary system is observed in a PN, relevance to PN evolution can be none (for very wide binaries) to a complicated history, with the symbiotic phenomenon being one of the possible outcomes.  

\section{Conclusions}

Recent observations have revealed PN structures that are much more complicated than a spherical shell.  Even a seemingly simple symmetric nebula such as the Ring Nebula in fact has a multiple bipolar outflow history. While we have learned that interacting winds  plus time-dependent photoionization as the result of stellar evolution can lead to multiple-shell structures \citep{sch05}, they can also be the result of projection of bipolar or multi-polar lobes plus torus.  Such confusion can only be sorted by kinematics.  The traditional long-slit spectroscopy does provide a certain amount of information, but we really need complete kinematic information of the entire nebula, including the shell, the crown, the lobes, and the halo.  For the ionized region, this can be achieved by
integral field spectroscopy and for the neutral envelope, by molecular line imaging.  Together with 3-D modeling (e.g., SHAPE), we may be able to derive the true  3-D structure.

An understanding of the real geometric structure of PNe is the first step toward understanding of their dynamical evolution.  This should begin with wide-field, high dynamic-range, multi-wavelength, imaging spectroscopic studies of the brightest PNe.  Only after we have gained a complete understanding of the 3-D structures of PNe such as the Ring, the Dumbbell, the Helix, would we be in a position to talk about the structure of other fainter PNe with confidence.

Our common perception of morphology of PNe is guided by their optical appearances.  Since ionized gas represents a small fraction of the total mass of PNe, optical structures cannot be used to infer the total mass distribution.
Regarding the physical cause of the bipolar/multipolar phenomena, we raise the possibility that this is not so much the result of dynamical ejection, but effects of illumination and ionization.  The bipolar lobes are in fact ionization-bounded low-density regions confined by high-density neutral gas.  This model can be tested by high angular resolution imaging of the molecular and dust components, a task that can be accomplished by {\it ALMA}.

\section*{Acknowledgments} 

I thank Nico Koning for construction of the models shown in Figures \ref{mycn18} and \ref{ngc7009} and Mei-Yan Wang for processing of images shown in Fig.~\ref{cfht1}.
The work was supported by a grant from the Research Grants Council of the Hong Kong Special Administrative Region, China (Project No. HKU 7028/07P).


\end{document}